\documentclass[aip,jcp,reprint,noshowkeys,superscriptaddress]{revtex4-1}
\usepackage{graphicx,dcolumn,bm,xcolor,microtype,multirow,amscd,amsmath,amssymb,amsfonts,physics,wrapfig,txfonts}
\usepackage[version=4]{mhchem}

\usepackage{siunitx}
\DeclareSIUnit[number-unit-product = {\,}]
\cal{cal}
\DeclareSIUnit\kcal{\kilo\cal}

\newcommand{\alert}[1]{\textcolor{black}{#1}}
\usepackage[normalem]{ulem}

\newcommand{\ie}{\textit{i.e.}}

\newcommand{\mc}{\multicolumn}

\newcommand{\QP}{\textsc{quantum package}}
\newcommand{\MRCC}{\textsc{mrcc}}
\newcommand{\CFOUR}{\textsc{cfour}}

\newcommand{\hH}{\Hat{H}}
\newcommand{\hT}{\Hat{T}}
\newcommand{\bH}{\Bar{H}}

\newcommand{\npis}{n \to \pi^*}
\newcommand{\pipis}{\pi \to \pi^*}

\usepackage[
	colorlinks=true,
    citecolor=blue,
    breaklinks=true
	]{hyperref}
\begin{document}

\newcommand{\LCPQ}{Laboratoire de Chimie et Physique Quantiques (UMR 5626), Universit\'e de Toulouse, CNRS, UPS, France}
\newcommand{\SMU}{Department of Chemistry, Southern Methodist University, Dallas, Texas 75275, USA}
\newcommand{\UOP}{Dipartimento di Chimica e Chimica Industriale, University of Pisa, Via Moruzzi 3, 56124 Pisa, Italy}
\newcommand{\CEISAM}{Universit\'e de Nantes, CNRS,  CEISAM UMR 6230, F-44000 Nantes, France}

\title{How accurate are EOM-CC4 vertical excitation energies?}

\author{Pierre-Fran\c{c}ois Loos}
	\email{loos@irsamc.ups-tlse.fr}
	\affiliation{\LCPQ}
\author{Devin A.~Matthews}
	\affiliation{\SMU}
\author{Filippo Lipparini}
	\affiliation{\UOP}
\author{Denis Jacquemin}
	\email{denis.jacquemin@univ-nantes.fr}
	\affiliation{\CEISAM}

\begin{abstract}
We report the first investigation of the performance of EOM-CC4 --- an approximate equation-of-motion coupled-cluster model which includes iterative quadruple excitations --- for vertical excitation energies in molecular systems.
By considering a set of 28 excited states in 10 small molecules for which we have computed CCSDTQP and FCI reference energies, we show that, in the case of excited states with a dominant contribution from the single excitations, CC4 yields excitation energies with sub-kJ~mol$^{-1}$ accuracy \alert{(\textit{i.e.}, error below $0.01$ eV)}, in very close agreement with its more expensive CCSDTQ parent.
Therefore, if one aims at high accuracy, CC4 stands as a highly competitive approximate method to model molecular excited states, with a significant improvement over both CC3 and CCSDT.
Our results also evidence that, although the same qualitative conclusions hold, one cannot reach the same level of accuracy for transitions \alert{with a dominant contribution from the double excitations}.
\end{abstract}

\maketitle

\section{Introduction}

Single-reference coupled-cluster (CC) theory provides a hierarchy of size-extensive methods delivering increasingly accurate energies and properties via the systematic increase of the maximum excitation degree of the cluster operator $\hT = \hT_1 + \hT_2 + \ldots + \hT_k$ (where $k \le n$ and $n$ is the number of electrons). \cite{Cizek_1966,Paldus_1972,Crawford_2000,Bartlett_2007,Shavitt_2009}
Without any truncation (\ie, $k = n$), the so-called full CC (FCC) method is equivalent to full configuration interaction (FCI), hence providing the exact energy and wave function of the system for a given atomic basis set.
However, it is not computationally viable due to its exponential scaling with system size, and one has to resort to truncated CC methods  (\ie, $k \ll n$) for computational convenience.
Popular choices are CC with singles and doubles (CCSD), \cite{Cizek_1966,Purvis_1982} CC with singles, doubles, and triples (CCSDT), \cite{Noga_1987a,Scuseria_1988} CC with singles, doubles, triples, and quadruples (CCSDTQ), \cite{Oliphant_1991,Kucharski_1992} and 
CC with singles, doubles, triples, quadruples, and pentuples (CCSDTQP) \cite{Hirata_2000,Kallay_2001} with corresponding computational scalings of $\order*{N^{6}}$, $\order*{N^{8}}$, $\order*{N^{10}}$, and  $\order*{N^{12}}$, respectively (where $N$ denotes the number of orbitals).
An alternative, systematically-improvable family of methods is defined by the CC2, \cite{Christiansen_1995a} CC3, \cite{Christiansen_1995b,Koch_1997} and CC4 \cite{Kallay_2005} series of models which have been introduced by the Aarhus group in the context of CC response theory. \cite{Christiansen_1998}
These iterative methods scale as $\order*{N^{5}}$, $\order*{N^{7}}$, and $\order*{N^{9}}$, respectively, and can be seen as cheaper approximations of CCSD, CCSDT, and CCSDTQ, by skipping the most expensive terms and avoiding the storage of the higher-excitation amplitudes.
A somewhat similar strategy has been applied to define the CCSDT-3 \cite{Urban_1985,Noga_1987b} and CCSDTQ-3 \cite{Kallay_2005} models based on arguments stemming from perturbation theory.
Of course, a large number of other approximate CC models have been developed over the years and we refer the interested reader to specialized reviews for more details. \cite{Crawford_2000,Piecuch_2002,Bartlett_2007,Shavitt_2009}

Coupled-cluster methods have been particularly successful for small- and medium-sized molecules in the field of thermodynamics, kinetics, and spectroscopy, thanks to the computations of accurate equilibrium geometries, \cite{Kallay_2003} potential energy surfaces, vibrational frequencies, \cite{Kallay_2004a} Born-Oppenheimer corrections, \cite{Gauss_2006} and a vast panel of properties such as dipoles (and higher moments), \cite{Kallay_2003}  NMR chemical shifts, \cite{Kallay_2004a} magnetizabilities, \cite{Gauss_2009} polarizabilities, \cite{Kallay_2006} etc.
Although originally developed for ground-state energies and properties, CC has been successfully extended to excited states \cite{Monkhorst_1977} thanks to the equation-of-motion (EOM) \cite{Emrich_1981,Comeau_1993,Stanton_1993,Krylov_2008} and linear-response (LR) \cite{Ghosh_1981,Sekino_1984,Koch_1990a,Koch_1990b,Rico_1993} formalisms which are known to produce identical excitation energies but different properties. 
In EOM-CC, one determines vertical excitation energies via the diagonalization of the similarity-transformed Hamiltonian $\bH = e^{-\hT} \hH e^{\hT}$.
A more general procedure to compute excitation energies that can be applied to any approximate CC model (as the ones mentioned above) consists in diagonalizing the so-called CC Jacobian obtained via the differentiation of the CC amplitude equations.
Again, by increasing the maximum excitation degree of $\hT$, one can systematically produce increasingly accurate EOM/LR excitation energies with the ``complete'' series CCSD, \cite{Stanton_1993} CCSDT, \cite{Kowalski_2001,Kucharski_2001} CCSDTQ, \cite{Hirata_2004} CCSDTQP\cite{Kallay_2004b} \ldots~or the ``approximate'' series CC2, \cite{Christiansen_1995a} CC3, \cite{Christiansen_1995b} CC4 \cite{Kallay_2004b} \ldots~while retaining the formal scaling of their ground-state analog. 
From here on, we drop the EOM prefix as all our calculations are performed within this approach.

Recently, a large collection of deterministic, stochastic or hybrid selected CI (SCI) methods \cite{Bender_1969,Whitten_1969,Huron_1973} has (re)appeared \cite{Booth_2009,Deustua_2017,Holmes_2016,Eriksen_2017,Xu_2018,Giner_2013,Giner_2015,Evangelista_2014,Tubman_2016,Liu_2016,Ohtsuka_2017,Zimmerman_2017,Coe_2018,Garniron_2018,Eriksen_2019} in the electronic structure landscape providing an alternative route to highly-accurate ground- and excited-state energies \cite{Holmes_2017,Li_2018,Li_2020,Loos_2018a,Chien_2018,Loos_2019,Loos_2020b,Loos_2020c,Loos_2020e,qp2,Eriksen_2020,Yao_2020,Veril_2021} (see Refs.~\onlinecite{Loos_2020a,Eriksen_2021} for recent reviews).
The general idea behind SCI methods is simple: rather than exploring the entire FCI space by systematically increasing the maximum excitation degree of the determinants taken into account (leading to the slowly convergent and size-inconsistent series of methods CISD, CISDT, CISDTQ, \ldots), one performs a sparse exploration of the FCI space by selecting only the most energetically important determinants thanks to a suitable criterion usually based on perturbation theory. \cite{Huron_1973,Garniron_2017,Sharma_2017,qp2}
By iteratively increasing the number of determinants of the variational space and supplementing it with a second-order perturbative correction (PT2), the SCI+PT2 family of methods has been recently shown to produce near-FCI correlation and excitation energies for small- and medium-size molecules in compact basis sets.\cite{Holmes_2017,Loos_2018a,Chien_2018,Loos_2019,Loos_2020b,Loos_2020c,qp2,Veril_2021}
Although the formal scaling of such algorithms remain exponential, the prefactor is greatly reduced which explains their current attractiveness in the electronic structure community and much wider applicability than their standard FCI parent. 

Taking advantage of the high accuracy of CC and SCI+PT2 methods, we have very recently created a large dataset gathering more than 500 highly-accurate vertical excitation energies for electronic transitions of various natures (valence, Rydberg, $\npis$, $\pipis$, singlet, doublet, triplet, charge-transfer, and double excitations) in small- and medium-sized molecules ranging from diatomics to molecules as large as naphthalene. \cite{Loos_2018a,Loos_2019,Loos_2020a,Loos_2020b,Loos_2020c,Veril_2021}
The main purpose of this so-called QUEST database is to provide reference excitation energies in order to perform fair and reliable benchmarks between electronic structure methods and assess their strengths and weakness for a large panel of chemical scenarios.
Most of these reference transition energies, which rely exclusively on high-level \textit{ab initio} calculations, can be reasonably considered as chemically accurate, \ie, within \SI{1}{\kcal\per\mol} or \SI{0.043}{\eV} of the FCI limit.
However, their accuracy may rapidly deteriorate, in particular, as the system size grows.
Indeed, it is usually challenging to compute reliable SCI+PT2 estimates or CCSDTQ excitation energies for molecules with more than four non-hydrogen atoms.
Therefore, for the larger molecules of the QUEST database, we mostly relied on CCSDT to define reference excitation energies.
Of course, it would be highly valuable to have access to more accurate methods (including, at least, quadruple excitations) in order to refine these theoretical best estimates.

In this context, the main purpose of the present study is to assess the relative accuracy of the approximate iterative CC4 model against the more expensive CCSDTQ and CCSDTQP methods in the case of vertical excitation energies, as well as their absolute accuracy with respect to FCI.
To do so, we consider a set of 10 small molecules (\ce{NH3}, \ce{C2}, \ce{BH}, \ce{BF}, \ce{CO}, \ce{N2}, \ce{HCl}, \ce{H2S}, \ce{HNO} and \ce{H2O}) and we compare the excitation energies associated with 28 singlet excited states of various natures ($\npis$, $\pipis$, Rydberg, valence, charge-transfer, and double excitations) and spatial symmetries obtained with various high-level CC methods.
Although a small number of studies have been published on the performance of CC4 for ground-state energies and properties, \cite{Kallay_2005,Matthews_2021} this work stands, to the best of our knowledge, as the first to consider CC4 for the computation of excited-states energies.
As we shall see below, CC4 is an excellent approximation to its CCSDTQ parent, and produces, in the case of excited states with a dominant contribution from the single excitations, excitation energies with sub-\si{\kJ\per\mol} accuracy \alert{(\textit{i.e.}, error below $0.01$ eV)} for this set of small molecules, well below the chemical accuracy threshold.

\section{Computational details}

\begin{table}
	\caption{Methods considered in the present study, their formal computational scaling and the electronic structure software employed to compute excitation energies.
	Here, $N$ is the number of orbitals.
	\label{tab:scaling}}
	\begin{ruledtabular}
	\begin{tabular}{lccc}
		Method	&	Scaling					&	Code		&	Ref.					\\
		\hline
		CC2			&	$\order*{N^{5}}$	&	\CFOUR		&	\onlinecite{cfour}		\\
		CCSD		&	$\order*{N^{6}}$	&	\CFOUR		&	\onlinecite{cfour}		\\
		CC3			&	$\order*{N^{7}}$	&	\CFOUR		&	\onlinecite{cfour}		\\
		CCSDT		&	$\order*{N^{8}}$	&	\CFOUR		&	\onlinecite{cfour}		\\
		CC4			&	$\order*{N^{9}}$	&	\CFOUR		&	\onlinecite{cfour}		\\
		CCSDTQ		&	$\order*{N^{10}}$	&	\CFOUR		&	\onlinecite{cfour}		\\
		CCSDTQP		&	$\order*{N^{12}}$	&	\MRCC		&	\onlinecite{mrcc}		\\	
		CIPSI		&	$\order*{e^{N}}$	&	\QP			&	\onlinecite{qp2}		\\
	\end{tabular}
	\end{ruledtabular}
\end{table}

All the methods considered in the present study are listed in Table \ref{tab:scaling} alongside their formal computational scaling and the electronic structure software employed to compute the excitation energies.
In a nutshell, we have used by default {\CFOUR} \cite{cfour} to compute the CC energies at the notable exception of CCSDTQP for which we have employed {\MRCC} and its automated implementation of high-order CC methods. \cite{mrcc}
The present {\CFOUR} calculations have been performed with the new and fast CC module (\texttt{xncc})  written by one of authors (DAM) which couples a general algebraic
and graphical interpretation of the non-orthogonal spin-adaptation approach with highly efficient storage format and implementation techniques designed to minimize data movement and to avoid costly tensor transposes. \cite{Matthews_2015b}
The FCI estimates were obtained with the SCI algorithm known as \textit{``configuration interaction using a perturbative selection made iteratively''} (CIPSI) implemented in {\QP}. \cite{qp2}
The error bars associated with the extrapolation step of the CIPSI calculations (see Ref.~\onlinecite{qp2}) have been computed using our recently developed protocol presented in Ref.~\onlinecite{Veril_2021}.

Because this is, to our knowledge, the first implementation of EOM-CC4, we have verified its accuracy by comparing the excitation energies obtained from solving for the right- and left-hand wave functions. 
Coupled cluster is a non-hermitian theory and thus the right- and left-hand eigenfunctions of the Jacobian are distinct, albeit with the same eigenvalue. 
For the left-hand solution, we have reused the already-verified code for the ground-state $\Lambda$	equations which describe the amplitude relaxation contribution in the analytic gradient theory. \cite{Matthews_2021} 
The structure of the left-hand EOM-CC and $\Lambda$ equations are identical, and so simply interfacing this existing code with a Davidson solver \cite{Davidson_1975} provides left-hand EOM-CC solutions; this procedure has been checked for other known-good methods such as CCSDT and CC3.

All calculations have been performed in the frozen-core approximation and the CC3/aug-cc-pVTZ geometries of the \alert{systems considered} here have been extracted from previous studies. \cite{Loos_2018a,Chrayteh_2021}
In the following, we consider diffuse-containing Dunning's double- and triple-$\zeta$ basis sets (aug-cc-pVDZ and aug-cc-pVTZ).
Note that CCSDTQP energies could only be computed for the smaller basis (aug-cc-pVDZ).

\section{Results and discussion}

\begin{squeezetable}
\begin{table*}
	\caption{Vertical excitation energies (in eV) of a selection of singly-excited states obtained at various levels of theory with the aug-cc-pVDZ and aug-cc-pVTZ basis sets. 
	$\%T_1$ is the percentage of single excitations involved in the transition computed at the CC3/aug-cc-pVTZ level.
	For the aug-cc-pVDZ basis, the mean absolute error (MAE), mean signed error (MSE), and maximum error (Max) with respect to CCSDTQP is reported.
	\alert{For the FCI data, the error bars reported in parenthesis correspond to one standard deviation.}
	\label{tab:singles}}
	\begin{ruledtabular}
	\begin{tabular}{llrrrrrrrrrrrrrrrr}
				&		&	\mc{8}{c}{aug-cc-pVDZ}		&		\mc{7}{c}{aug-cc-pVTZ}		\\	
				\cline{4-11} \cline{12-18}
	Mol.	&	State	& $\%T_1$	&CC2	&CCSD	&CC3	&CCSDT	&CC4	&CCSDTQ	&CCSDTQP	&FCI	
									&CC2	&CCSD	&CC3	&CCSDT	&CC4	&CCSDTQ	&FCI			\\
	\hline
	\ce{NH3}	&	$^1A_2$ 		&93	&6.249	&6.455	&6.464	&6.462	&6.479	&6.480	&6.482	&6.483(1)	&6.387	&6.600	&6.573	&6.571	&6.585	&6.586	&6.593(22)	\\	
				&	$^1E$			&93	&7.733	&8.024	&8.061	&8.057	&8.078	&8.079	&8.081	&8.082(1)	&7.847	&8.148	&8.146	&8.143	&8.161	&8.161	&8.171(20)	\\	
				&	$^1A_1$ 		&94	&9.400	&9.649	&9.664	&9.659	&9.677	&9.677	&9.680	&9.681(8)	&9.051	&9.334	&9.318	&9.314	&9.331	&9.331	&9.340(19)	\\
				&	$^1A_2$ 		&93	&10.148	&10.376	&10.396	&10.391	&10.409	&10.409	&10.411	&10.412(1)	&9.654	&9.953	&9.945	&9.939	&9.957	&9.957	&9.967(19)	\\
	\ce{BH}		&	$^1\Pi$ 		&95	&2.862	&2.970	&2.955	&2.946	&2.947	&2.947	&2.947	&2.947(0)	&2.831	&2.928	&2.910	&2.900	&2.901	&2.901	&2.901(0)	\\
	\ce{BF}		&	$^1\Pi$ 		&94	&6.509	&6.534	&6.478	&6.491	&6.484	&6.485	&6.485	&6.485(1)	&6.445	&6.464	&6.410	&6.423	&6.416	&6.417	&6.418(2)\\
	\ce{CO}		&	$^1\Pi$ 		&93	&8.724	&8.671	&8.572	&8.574	&8.562	&8.563	&8.561	&8.563(4)	&8.638	&8.587	&8.486	&8.492	&8.479	&8.480	&	\\
				&	$^1\Sigma^-$ 	&93	&10.381	&10.096	&10.122	&10.062	&10.055	&10.057	&10.057	&10.056(1)	&10.297	&9.986	&9.992	&9.940	&9.930	&9.932	&	\\
				&	$^1\Delta$ 		&91	&10.685	&10.210	&10.225	&10.178	&10.167	&10.169	&10.168	&10.168(1)	&10.604	&10.123	&10.119	&10.076	&10.064	&10.066	&	\\
				&	$^1\Sigma^+$ 	&91	&11.089	&11.171	&10.917	&10.944	&10.925	&10.926	&10.919	&			&11.106	&11.222	&10.943	&10.987	&10.961	&10.963	&	\\
				&	$^1\Sigma^+$ 	&92	&11.628	&11.710	&11.483	&11.518	&11.510	&11.510	&11.506	&			&11.626	&11.751	&11.489	&11.540	&11.521	&11.523	&	\\
				&	$^1\Pi$ 		&92	&11.878	&11.973	&11.737	&11.767	&11.757	&11.758	&11.753	&			&11.825	&11.960	&11.690	&11.737	&11.719	&11.720	&	\\
	\ce{N2}		&	$^1\Pi_g$  		&92	&9,528	&9.495	&9.442	&9.417	&9.409	&9.411	&9.409	&9.411(3)	&9.439	&9.408	&9.344	&9.326	&9.317	&9.319	&	\\
				&	$^1\Sigma_u^-$	&97	&10.428	&10.197	&10.059	&10.060	&10.063	&10.055	&10.054	&10.054(0)	&10.320	&9.996	&9.885	&9.890	&9.883	&9.878	&9.879(4)\\
				&	$^1\Delta_u$ 	&95	&10.961	&10.607	&10.433	&10.436	&10.439	&10.430	&10.428	&10.429(0)	&10.863	&10.443	&10.293	&10.302	&10.294	&10.287	&10.289(12)\\
				&	$^1\Sigma_g^+$	&92	&13.077	&13.326	&13.229	&13.202	&13.171	&13.182	&13.181	&13.180(1)	&12.833	&13.151	&13.013	&12.999	&12.962	&12.974	&	\\
				&	$^1\Pi_u$  		&82	&13.309	&13.451	&13.279	&13.174	&13.128	&13.131	&13.127	&			&13.152	&13.422	&13.223	&13.140	&13.091	&13,095	&	\\
				&	$^1\Sigma_u^+$	&92	&12.937	&13.250	&13.146	&13.130	&13.099	&13.109	&13.107	&			&12.888	&13.263	&13.120	&13.118	&13.078	&13.090	&	\\
				&	$^1\Pi_u$ 		&87	&14.091	&13.765	&13.635	&13.591	&13.551	&13.560	&13.558	&			&13.963	&13.674	&13.494	&13.455	&13.409	&13.419\\
	\ce{HCl}	&	$^1\Pi$ 		&94	&7.895	&7.862	&7.819	&7.815	&7.822	&7.822	&7.823	&7.823(0)	&7.959	&7.906	&7.840	&7.834	&7.837	&7.837	&7.838(1)\\
	\ce{H2S}	&	$^1B_1$  		&94	&6.157	&6.141	&6.098	&6.098	&6.102	&6.103	&6.103	&6.103(1)	&6.304	&6.294	&6.240	&6.237	&6.238	&6.238	&6.240(7)\\
				&	$^1A_2$  		&94	&6.431	&6.343	&6.293	&6.286	&6.286	&6.286	&6.286	&6.286(0)	&6.345	&6.246	&6.192	&6.185	&6.181	&6.181	&6.181(6) \\
	\ce{H2O}	&	$^1B_1$  		&93	&7.089	&7.447	&7.511	&7.497	&7.531	&7.528	&7.532	&7.533(0)	&7.234	&7.597	&7.605	&7.591	&7.623	&7.620	&7.626(3)\\
				&	$^1A_2$  		&93	&8.743	&9.213	&9.293	&9.279	&9.317	&9.313	&9.318	&9.318(0)	&8.889	&9.361	&9.382	&9.368	&9.405	&9.400	&9.407(7)\\
				&	$^1A_1$  		&93	&9.486	&9.861	&9.921	&9.903	&9.940	&9.937	&9.941	&9.941(0)	&9.580	&9.957	&9.966	&9.949	&9.986	&9.981	&9.986(2)\\
	\hline
	MAE			&					&	&0.257	&0.109	&0.028	&0.018	&0.003	&0.002	\\
	MSE			&					&	&0.020	&0.075	&0.013	&0.000	&0.000	&0.000	\\
	Max			&					&	&0.575	&0.324	&0.152	&0.047	&0.011	&0.007	\\
	\end{tabular}
	\end{ruledtabular}
\end{table*}
\end{squeezetable}

\begin{figure*}
	\includegraphics[width=0.7\linewidth]{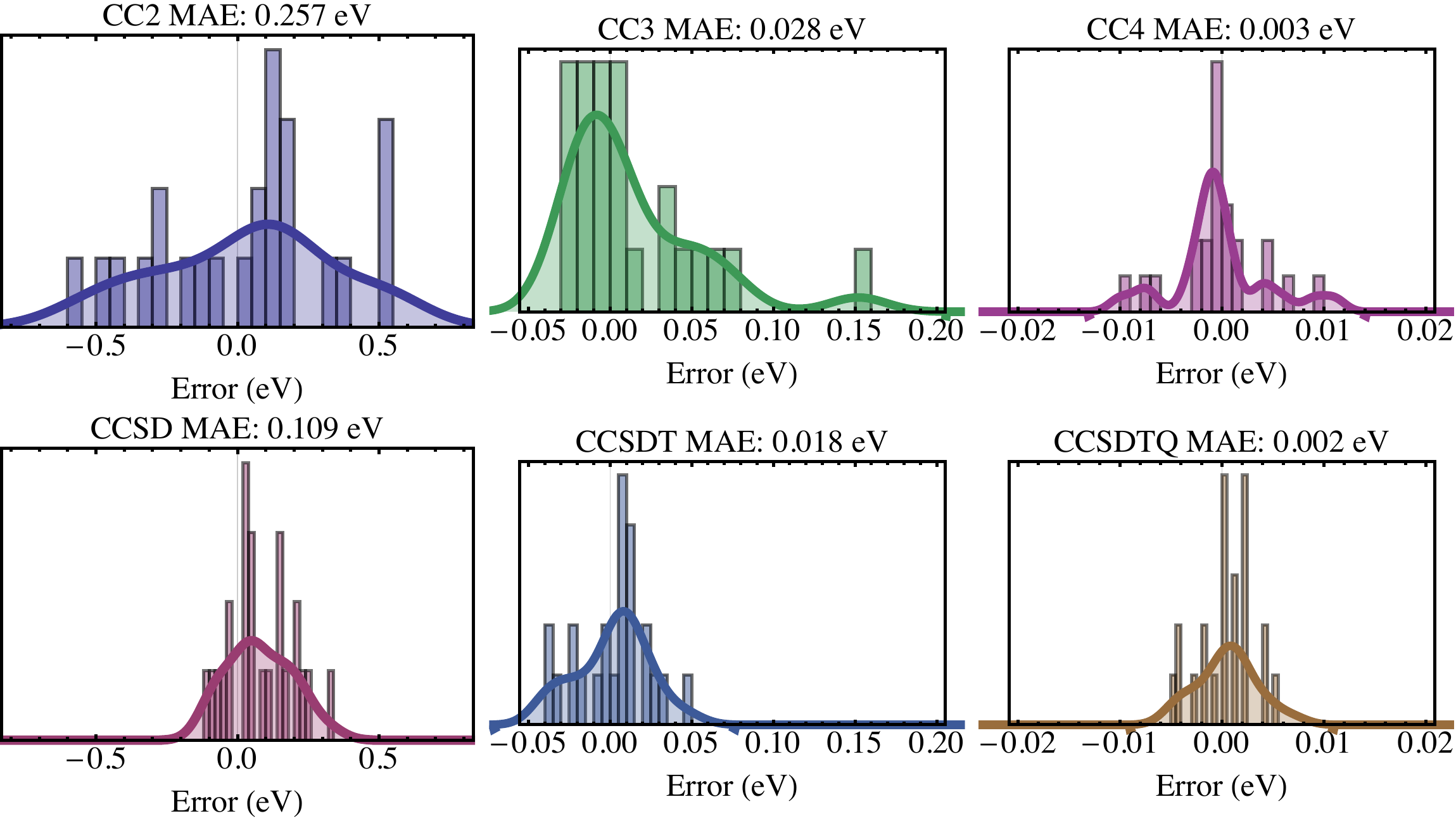}
	\caption{
	Distribution of the error (in eV) in excitation energies (with respect to CCSDTQP) for CC2, CC3, and CC4 (top) and CCSD, CCSDT, and CCSDTQ (bottom) obtained with the aug-cc-pVDZ basis.
	Note the various error ranges ($X$ axis) for the various methods.
	See Table \ref{tab:singles} for the raw data.
	\label{fig:error}}
\end{figure*}

Table \ref{tab:singles} gathers, for the two considered basis sets, 25 vertical excitation energies with a (strongly) dominant contribution from the single excitations computed for a set of 8 molecules with various CC models as well as the FCI estimates computed with CIPSI.
First, we underline that the FCI estimates show how accurate the CCSDTQP reference data are, with a maximum deviation of \SI{0.002}{\eV} when one considers the aug-cc-pVDZ basis. 
For the larger aug-cc-pVTZ basis, the CCSDTQ and FCI remain in excellent agreement although the error bars associated with the extrapolated FCI values prevent us from any quantitative comparisons.

The mean absolute errors (MAEs) and mean signed errors (MSEs) with respect to CCSDTQP  computed in the aug-cc-pVDZ basis are reported in the bottom of Table \ref{tab:singles} for the CC2-to-CCSDTQ models.
The distribution of the errors are reported in Fig.~\ref{fig:error} for each level of theory.
These statistical quantities nicely illustrate the systematic improvement of the transition energies when one ramps up the computational effort following the series CC2, CCSD, CC3, CCSDT, CC4, and CCSDTQ.
We note that the errors decrease by roughly one order of magnitude when switching from CCSD to CC3 and from CCSDT to CC4, whilst improvements of approximately 50\% ``only'' are noted when going from the ``approximate'' model to the ``complete'' method (\ie, from CC2 to CCSD, from CC3 to CCSDT, and from CC4 to CCSDTQ).
In other words, CC4 brings significant improvements in terms of MAE and MSE as compared to the third-order methods, CC3 and CCSDT, which demonstrates the importance of quadruple excitations when one aims at very high accuracy.  
Besides, for the two basis sets, there is an outstanding similarity between the CC4 and CCSDTQ excitation energies with mean absolute and signed deviations below (equal to) \SI{0.001}{\eV} and a maximum deviation of \SI{0.011}{\eV} (\SI{0.007}{\eV}) between to the two sets of data obtained with the aug-cc-pVDZ (aug-cc-pVTZ) basis set. 
Therefore, including quadruples allows to reach sub-\si{\kJ\per\mol} accuracy (\ie, average error below \SI{0.01}{\eV}) for singly-excited states with only a rather minor improvement in going from CC4 to CCSDTQ.

\begin{squeezetable}
\begin{table}
	\caption{Vertical excitation energies (in eV) for a selection of \alert{transitions with a dominant contribution from the double excitations} obtained at various levels of theory with the aug-cc-pVDZ basis set. 
	$\%T_1$ is the percentage of single excitations involved in the transition (computed at the CC3/aug-cc-pVTZ level).
	\label{tab:doubles}}
	\begin{ruledtabular}
	\begin{tabular}{llrrrrrrr}
	Mol.		&	State			& $\%T_1$	&CC3	&CCSDT	&CC4	&CCSDTQ	&CCSDTQP	&FCI	\\
	\hline
	\ce{C2}		&	$^1\Delta_g$ 	&	1		&3.107	&2.632	&2.341	&2.241	&2.214	&2.213(0)	\\	
				&	$^1\Sigma_g^+$ &	1		&3.283	&2.874	&2.602	&2.521	&2.505	&2.503(1)	\\	
	\ce{HNO}	&	$^1A'$ 			&	0		&5.247	&4.756	&4.454	&4.424	&		&4.397(1)	\\
	\end{tabular}
	\end{ruledtabular}
\end{table}
\end{squeezetable}

\begin{figure*}
	\includegraphics[width=0.7\linewidth]{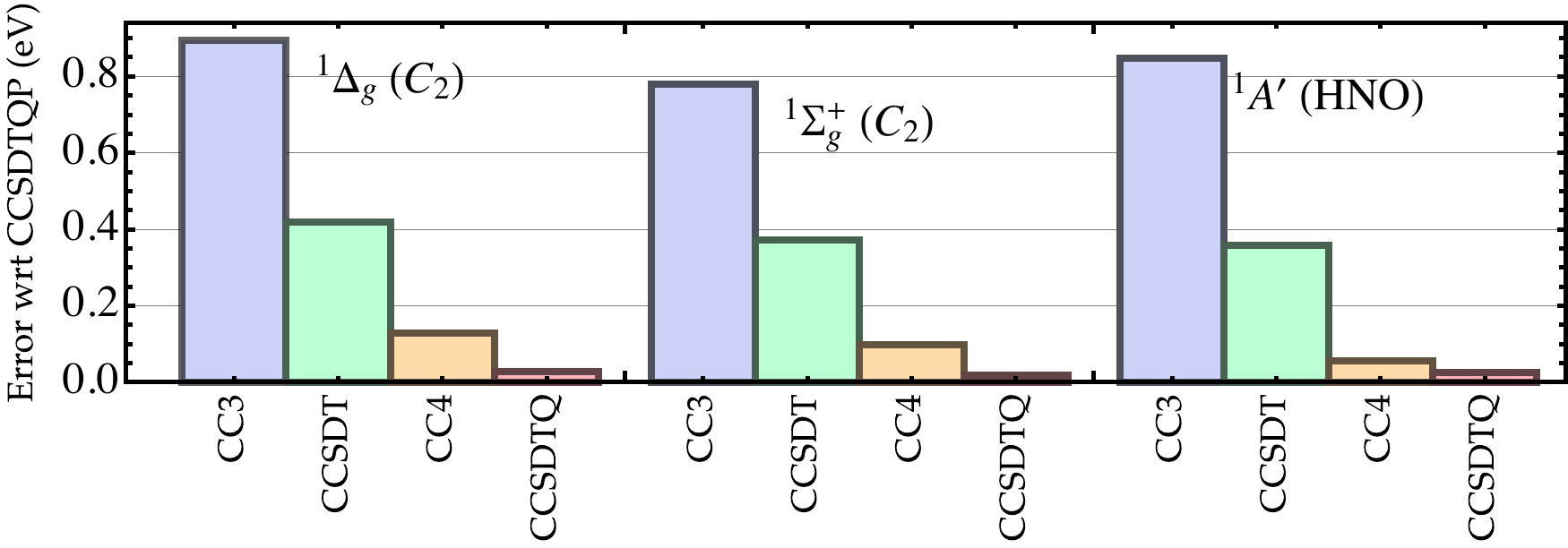}
	\caption{
	Error in excitation energies (with respect to FCI) computed with CC3, CCSDT, CC4, and CCSDTQ in the aug-cc-pVDZ basis
	for \alert{transitions with a dominant contribution from the double excitations} gathered in Table \ref{tab:doubles}.
	\label{fig:doubles}}
\end{figure*}

A closer inspection at Table \ref{tab:singles} shows that the largest deviations appear for the transitions with the smallest $\%T_1$ values (where $\%T_1$ is the percentage of single excitations involved in the transition which is computed at the CC3/aug-cc-pVTZ level in the present case). 
This is particularly noticeable for the two $^1\Pi_u$ transitions of \ce{N2}.
This suggests that the performance of the various CC models discussed above might be highly dependent on the nature of the transitions, as discussed in Ref.~\onlinecite{Loos_2019}.
To investigate further this point, we report in Table \ref{tab:doubles} vertical excitation energies for \alert{transitions with a dominant contribution from the doubly-excited determinants} in the carbon dimer, \ce{C2}, and nitroxyl, \ce{HNO}, computed with the aug-cc-pVDZ basis for the methods including at least triple excitations; the second-order methods, CC2 and CCSD, being unable to faithfully locate these \alert{states with a large contribution of double excitations.}
These transitions can be labeled as ``pure'' double excitations as they involve an insignificant amount of single excitations ($\%T_1 \approx 0$), hence providing a very stringent test for the EOM-CC formalism.
In this case, as shown in Fig.~\ref{fig:doubles}, the differences between methods are magnified, but the conclusions drawn in the previous paragraph hold: CC4 is an excellent approximation to CCSDTQ (with a maximum deviation of \SI{0.1}{\eV} for the $^1\Delta_g$ transition of \ce{C2}) and a massive improvement over CC3 (where the error can be as large as \SI{0.9}{\eV}) and, to a lesser extent, over CCSDT.
However, for these transitions with a dominant double excitation character, CC4 does not permit to reach chemical accuracy with errors of the order of \SI{0.1}{\eV} compared to CCSDTQP and FCI.
The outcome might differ for transitions of mixed characters ($\%T_1 \approx 70$) such as the well-known $^1 A_g$ excited state of butadiene.\cite{Maitra_2004,Saha_2006,Watson_2012,Shu_2017,Barca_2018a,Loos_2019} 

\section{Conclusion}

Thanks to the results gathered in the present study, we can conclude, for this set of small molecules at least, that CC4 is a rather competitive approximation to its more expensive CCSDTQ parent as well as a very significant improvement over both its third-order version, CC3, and the ``complete'' CCSDT method.
This is particularly true in the case of \alert{transitions with a dominant contribution from the single excitations} (Table \ref{tab:singles}) when one reaches sub-kJ~mol$^{-1}$ accuracy.
\alert{For states with a dominant contribution from the double excitations}, we have seen (Table \ref{tab:doubles}) that the same qualitative conclusions hold but one cannot reach chemical accuracy for the set of ``pure'' double excitations that we have considered.

These findings are promising, though we are well aware that the conclusions obtained for small and larger molecules might differ significantly.
For example, CCSD outperforms CC2 for compact molecules (as here), but the opposite trend is often found for larger compounds.\cite{Loos_2018a,Loos_2020a,Veril_2021} 
Therefore, although further investigations on larger compounds are definitely required, the present results are very encouraging as CC4, with its $\order*{N^9}$ scaling, can be applied to significantly larger molecules than CCSDTQ [which scales as $\order*{N^{10}}$]. 
This will likely allow us to revisit, in the future, some of the theoretical best estimates defined in the QUEST database.\cite{Loos_2020a,Veril_2021}

\begin{acknowledgements}
PFL has received funding from the European Research Council (ERC) under the European Union's Horizon 2020 research and innovation programme (Grant agreement No.~863481).
DAM has received funding from the National Science Foundation (grant No.~OAC-2003931).
This work was performed using HPC resources from CALMIP (Toulouse) under allocation 2021-18005 and of the CCIPL computational center installed in Nantes.
\end{acknowledgements}

\section*{Data availability statement}
The data that support the findings of this study are openly available in Zenodo at \href{http://doi.org/10.5281/zenodo.4739288}{http://doi.org/10.5281/zenodo.4739288}.

\section*{References}
\bibliography{CC4}

\end{document}